\documentclass[11pt,showpacs,preprintnumbers,amsmath,amssymb]{revtex4}
\usepackage{graphicx}
\begin{document}

\title{Phase Diagram of the Two-Dimensional Ising Model with Random
Competing Interactions}

\vskip \baselineskip

\author{Octavio R. Salmon$^{1}$}
\thanks{E-mail address: octavior@cbpf.br}

\author{J. Ricardo de Sousa$^{2,3}$}
\thanks{E-mail address: jsousa@pq.cnpq.br}

\author{Fernando D. Nobre$^{1,3}$}
\thanks{Corresponding author: E-mail address: fdnobre@cbpf.br}
\thanks{FAX: 55-21-21417401}
\thanks{Telephone: 55-21-21417513}

\address{
$^{1}$Centro Brasileiro de Pesquisas F\'{\i}sicas \\
Rua Xavier Sigaud 150 \\
22290-180 \hspace{5mm} Rio de Janeiro - RJ \hspace{5mm} Brazil \\
$^{2}$ Departamento de F\'{\i}sica \\ 
Universidade Federal do Amazonas \hspace{5mm} 3000 Japiim \\ 
69077-000 \hspace{5mm} Manaus-AM
\hspace{5mm} Brazil \\
$^{3}$National Institute of Science and Technology for Complex Systems \\ 
Rua Xavier Sigaud 150 \\
22290-180 \hspace{5mm} Rio de Janeiro - RJ \hspace{5mm} Brazil}

\date{\today}

\vspace{1cm}

\begin{abstract}

\begin{center}

{\large\bf Abstract}

\end{center}

\noindent
An Ising model with ferromagnetic
nearest-neighbor interactions $J_{1}$ ($J_{1}>0$) and random  
next-nearest-neighbor 
interactions [$+J_{2}$ with probability $p$ and $-J_{2}$ with 
probability $(1-p)$; $J_{2}>0$] is studied within the framework of an
effective-field theory based on the differential-operator technique.
The order parameters are calculated, considering finite clusters
with $n=1,2, {\rm and} \ 4$ spins, using the standard approximation 
of neglecting correlations. 
A phase diagram is obtained in the plane temperature versus
$p$, for the particular case $J_{1}=J_{2}$, showing both 
superantiferromagnetic (low $p$) and ferromagnetic (higher
values of $p$) orderings at low temperatures.  

\vspace{1cm}


\end{abstract}
\maketitle

\newpage

Many magnetic compounds are well-described in terms of theoretical models
characterized by a competition of nearest-neighbor and
next-nearest-neighbor interactions. As typical examples, one has 
${\rm Eu_{x}Sr_{1-x}S}$ \cite{maletta82,maletta83} and 
${\rm Fe_{x}Zn_{1-x}F_{2}}$ \cite{belangerreview}, which may present 
various low-temperature magnetic orderings, depending on its parameters, 
like the strength of these interactions and the concentration
of magnetic ions x. 
The simplest model for such systems consists of an Ising model 
with competing uniform 
interactions, ferromagnetic $J_{1}$ (or antiferromagnetic $-J_{1}$) 
($J_{1}>0$) on nearest-neighbor, and antiferromagnetic 
$-J_{2}$ ($J_{2}>0$) on 
next-nearest-neighbor spins (to be denoted hereafter as 
$J_{1}-J_{2}$ Ising model). 

The $J_{1}-J_{2}$ Ising model on a square
lattice has attracted the attention of many workers, being investigated 
through several approximation methods 
\cite{katsura,barber,wu,oitmaa,swendsen,landau80,binderlandau80,%
landaubinder85,moranlopez93,%
moranlopez94,buzano,tanaka,plascak,oliveira,blote79,nightingale77,%
blote85,slotte,zandvliet,aguilera,malakis06,monroekim,kalz,rosana,ohare}. 
The various possible phases are the paramagnetic ({\bf P}), for high
temperatures, whereas for low temperatures one may have the ferromagnetic
({\bf F}), antiferromagnetic ({\bf AF}) and superantiferromagnetic 
({\bf SAF}) phases; the later is characterized by
alternate ferromagnetic rows (or columns) of oppositely oriented spins. 
In the absence of a magnetic field, it can 
be shown that one has a symmetry with respect to the sign of the
nearest-neighbor interactions,  
i.e., the ferromagnetic and antiferromagnetic states are equivalent
\cite{barber}. 
At zero temperature (ground state), one has two ordered 
states depending on the value of the frustration parameter 
$\alpha=(J_{2}/J_{1})$, namely, the {\bf F} ($0<\alpha <1/2$)
(in an equivalent way, 
an {\bf AF} state occurs instead of the {\bf F} one, if the sign of
nearest-neighbor interactions is changed) 
and the {\bf SAF} ($\alpha >1/2$) phases.
On the other hand, its phase diagram for finite temperatures has 
been the object of some controversies. 
A continuous critical
frontier between the {\bf P} and {\bf F} phases, 
for $0 < \alpha <1/2$, with the critical temperature 
$T_{c}(\alpha)$ decreasing and approaching zero for 
$\alpha \rightarrow 1/2$, is well-accepted nowadays. 
However, some characteristics of this phase diagram for $\alpha>1/2$ are
polemic; in particular, the presence
of a first-order transition between the {\bf P} and {\bf SAF} phases, as
well as a tricritical point characterized by the 
coordinates ($\alpha _{t},T_{t}$), beyond which a continuous critical
frontier occurs, has been the object of some debate. Most of the works
indicate that within the range
$1/2 < \alpha < \alpha_{t}$ one should get a first-order phase transition
between the {\bf P} and {\bf SAF} phases, whereas for 
$\alpha>\alpha_{t}$ this transition should become continuous. 
Some studies using
cluster-variation methods~\cite{moranlopez93,moranlopez94,buzano}
claim a first-order line in the range $0.5<\alpha<1.2$, i.e., 
$\alpha_{t}=1.2$.
However, recent results of Monte Carlo simulations~\cite{malakis06,kalz},
analysis of zeros of the partition function~\cite{monroekim}, and
effective-field theory (EFT)~\cite{rosana} suggest 
that $\alpha_{t}<1$, in such a way that a continuous transition 
between the {\bf P} and {\bf SAF} phases should occur for $\alpha=1$.

In the present work we introduce a disordered version of the 
$J_{1}-J_{2}$ Ising model on a
square lattice. It is important to mention that due to the symmetry 
with respect to the sign of the 
nearest-neighbor interactions, for which a change of sign takes 
{\bf F} $\leftrightarrow$ {\bf AF}, a disorder that produces a simple
change of sign in $J_{1}$ is not expected to bring any novel physical
properties \cite{barber}. A disordered version of the 
$J_{1}-J_{2}$ Ising model has already been defined 
in Refs.~\cite{fytas08a,fytas08b} by considering the same bond 
disorder, $P(J_{ij})=(1/2)[\delta (J_{ij}-J_{1})+\delta(J_{ij}-J_{2})]$, 
for both nearest- and next-nearest-neighbor pairs of spins, 
which were introduced 
with the same sign in the corresponding Hamiltonian. In this analysis, the
authors have investigated particular choices for  
the ratio $J_{2}/J_{1}$, through numerical simulations of the model on a
square lattice, and have found as its main results: 
(i) An interesting saturation, with respect to the size of the system, of
the specific-heat maxima; (ii) Critical exponents following the
weak-universality scenario, according to which critical exponents change
with the disorder in a certain way to keep some ratios (e.g., $\beta/\nu$
and $\gamma/\nu$) unchanged  
at their corresponding pure values, similarly to what happens for other
disordered magnetic models.  
Herein, we introduce a different disordered version of the $J_{1}-J_{2}$
Ising model, by defining a disorder  
only in the next-nearest-neighbor interactions, in such a way that we shall
consider the Hamiltonian, 

\begin{equation} \label{1}
\mathcal{H}=-J_{1} \sum_{\text{nn}}\sigma _{i}
\sigma_{j} - \sum_{\text{nnn}}J_{ij}^{(2)}\sigma _{i}\sigma _{j} 
\quad (\sigma_{i}=\pm 1)~,  
\end{equation}

\vskip \baselineskip
\noindent
where the summations are over nearest-neighbor (nn) and
next-nearest-neighbor (nnn) pairs of spins, with 
ferromagnetic ($J_{1}>0$) and random $J_{ij}^{(2)}$
interactions, respectively. The later follow a 
bimodal probability distribution,

\begin{equation} \label{2}
\mathcal{P}(J_{ij}^{(2)})=p\delta (J_{ij}^{(2)}-J_{2})+
(1-p)\delta(J_{ij}^{(2)}+J_{2}) \quad (J_{2}>0).  
\end{equation}

\vskip \baselineskip
\noindent
The particular case $p=0$ of this model 
corresponds to the pure $J_{1}-J_{2}$ Ising model, described above and 
has been studied in the literature by many authors
\cite{katsura,barber,wu,oitmaa,swendsen,landau80,binderlandau80,%
landaubinder85,moranlopez93,%
moranlopez94,buzano,tanaka,plascak,oliveira,blote79,nightingale77,%
blote85,slotte,zandvliet,aguilera,malakis06,monroekim,kalz,rosana,ohare}. 
The introduction of the above type of disorder allows for the 
possibility of a
richer critical behavior, characterized by a competition between 
different types of orderings as one varies $p$, for a fixed ratio of the
interaction strengths, $\alpha=(J_{2}/J_{1})$. In addition to that, 
from the point of view of
physical realizations, the parameter $p$, which is usually associated with
the concentration of a given chemical element, is much easier to be varied
experimentally than the
ratio of interaction strengths, $\alpha$. 
In the present analysis we restrict ourselves to the case
$J_{2}=J_{1}$, i.e., $\alpha=1$. 

Herein, we treat this model by using the EFT method, which has been shown
recently to be very useful in the pure case (i.e., $p=0$) \cite{rosana}. 
The EFT technique is based on
rigorous correlation identities, as a starting point, and applies the
differential-operator technique developed by 
Honmura and Kaneyoshi \cite{honmura}. Within this method, the 
averages of a general function $\mathcal{A}(\{n\})$, involving spin variables, 
is obtained by

\begin{equation} \label{3}
\left\langle \mathcal{A}(\{n\})\right\rangle =\left\langle \frac{Tr_{\{n\}}%
\mathcal{A}(\{n\})e^{-\beta \mathcal{H}_{\{n\}}}}{Tr_{\{n\}}e^{-\beta 
\mathcal{H}_{\{n\}}}}\right\rangle~,  
\end{equation}

\begin{figure}[t]
\begin{center}
\includegraphics[width=0.45\textwidth,angle=0]{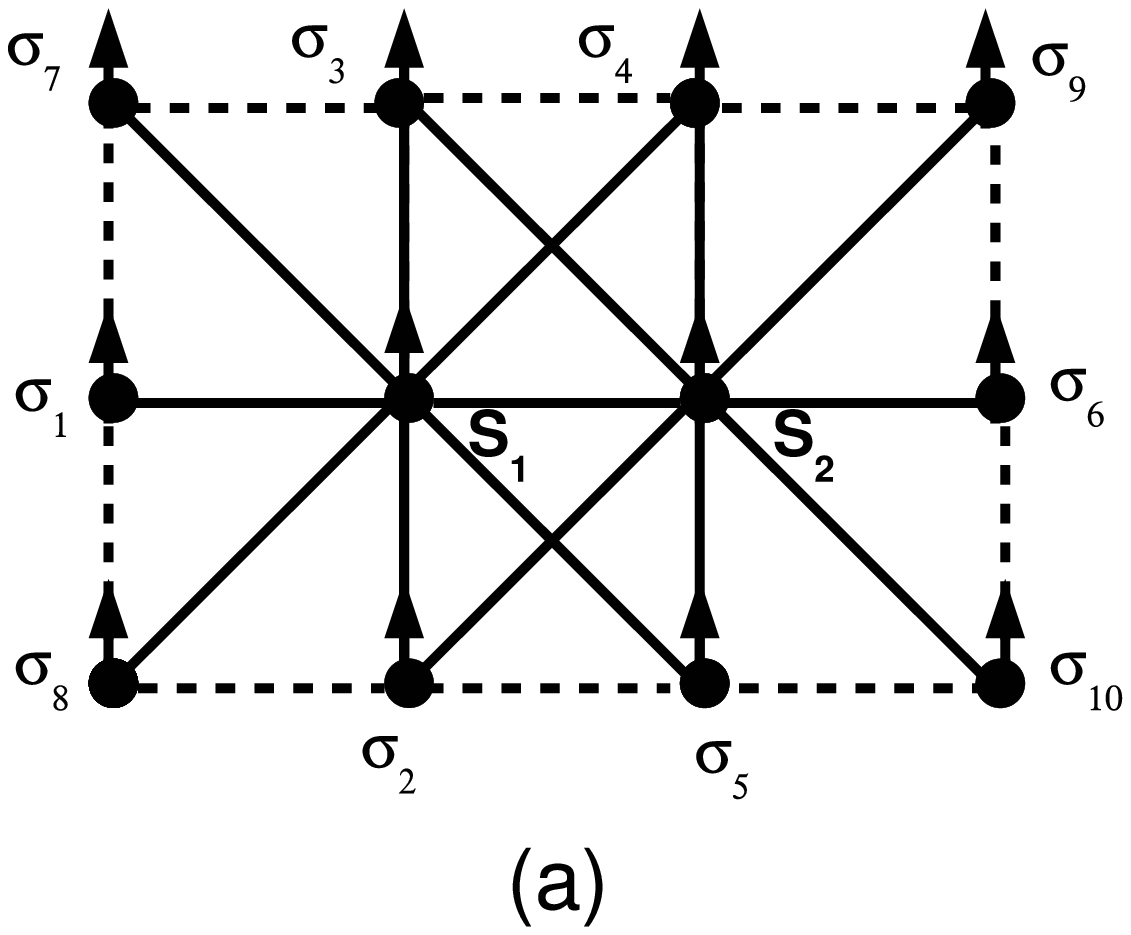}
\hspace{1.0cm}
\includegraphics[width=0.45\textwidth,angle=0]{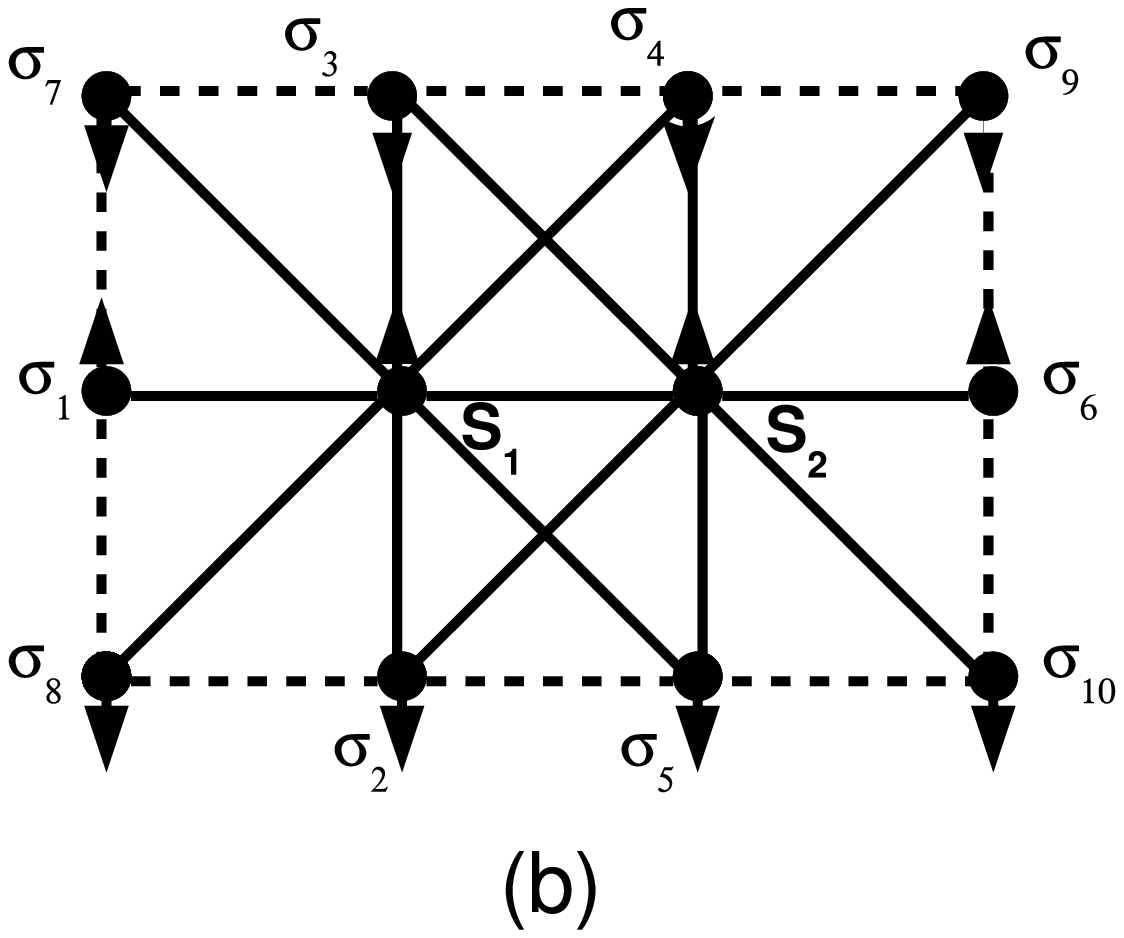}
\end{center}
\protect\caption{A cluster with $n=2$ (represented by spins   
$S_{1}$ and $S_{2}$) and its surrounding (spins 
$\sigma_{1},\sigma_{2}, \cdots, \sigma_{10}$), for the model defined in  
Eq.~(\ref{1}), in its ferromagnetic (a) and
superantiferromagnetic (b) ground states. The full lines represent the
interactions of spins $S_{1}$ and $S_{2}$, which are taken into account 
exactly in
Eq.~(\ref{3}), whereas the dashed lines represent the usual square-lattice
structure.} 
\label{fig1}
\end{figure}

\vskip \baselineskip
\noindent
where the partial trace $Tr_{\{n\}}$ is calculated exactly over the 
set $\{n\}$ of spin
variables that belong to the finite cluster specified by the multisite 
spin Hamiltonian 
$\mathcal{H}_{\{n\}}$ and 
$\left\langle \cdot \cdot \cdot \right\rangle $
indicates the usual canonical thermal average over the surrounding 
of the cluster. In the case $n=2$ (see Fig.~\ref{fig1}) the
trace $Tr_{\{n\}}$ is calculated over spins 
$S_{1}$ and $S_{2}$, whereas 
$\left\langle \cdot \cdot \cdot \right\rangle $ is to be considered over  
$\sigma_{1},\sigma_{2}, \cdots, \sigma_{10}$. 

We have applied the EFT
method for clusters of with $n=1,2, {\rm and} \ 4$ spins (to be denoted
hereafter EFT-$n$) and below we describe the case of EFT-2, illustrated in
Fig.~\ref{fig1}. The Hamiltonian for this cluster is given by 

\begin{equation} \label{4}
\mathcal{H}_{2}=-J_{1}S_{1}\cdot S_{2}+S_{1}a_{1}+S_{2}a_{2}~, 
\end{equation}

\noindent
with

\begin{equation} \label{5}
a_{1}=-J_{1}\sum\limits_{i=1}^{3}\sigma_{i}
\ -\sum_{j=4,5,7,8}J_{1j}^{(2)}\sigma_{j}~,
\end{equation}

\noindent
and

\begin{equation} \label{6}
a_{2}=-J_{1}\sum\limits_{i=4}^{6}\sigma_{i}
\ -\sum_{j=2,3,9,10}J_{2j}^{(2)}\sigma_{j}~. 
\end{equation}

\vskip \baselineskip
\noindent
We have divided the square lattice into two sublattices $A$ and $B$, 
each of them defined by alternating lines (or columns) in such a way that 
in Fig.~\ref{fig1} one has 
$m_{A}=\left\langle \left\langle S_{1}\right\rangle \right\rangle
_{J^{(2)}}=\left\langle \left\langle S_{2}\right\rangle \right\rangle
_{J^{(2)}}=\left\langle \left\langle \sigma _{i}\right\rangle \right\rangle 
_{J^{(2)}}$ 
($i=1,6$) and $m_{B}=\left\langle \left\langle \sigma _{i}\right\rangle
\right\rangle _{J^{(2)}}$ ($i=2,3,4,5,7,8,9,10$), where 
$\left\langle \cdot \cdot \right\rangle _{J^{(2)}}$ denotes an average 
over the disorder; in the {\bf F} ({\bf SAF}) phase one has that 
$m_{A}=m_{B}=m$ ($m_{A}=-m_{B}=m$).  
At zero temperature ($m=1$), the ground state of the Hamiltonian of 
Eq.~(\ref{1}) 
is ferromagnetic for $p=1$
[see Fig.~\ref{fig1}(a)], whereas for $p=0$ and $\alpha=1$ one gets the
superantiferromagnetic state [shown in Fig.~\ref{fig1}(b)]. 
In each case, a continuous phase transition occurs
at a finite temperature between the corresponding low-temperature state 
and the paramagnetic state. 
To the best of our knowledge,
theoretical works to investigate the critical behavior of this model for
$0<p<1$ have never been carried and this represents the 
purpose of the present work.

The parameter $m$ may be obtained by evaluating the 
average magnetization per spin in sublattice A,  
$m_{A}=\left\langle \left\langle \frac{1}{2}(S_{1}+S_{2})\right\rangle
\right\rangle _{J^{(2)}}$, calculating the inner trace in  
Eq.~(\ref{3}) over spins $S_{1},S_{2}=\pm 1$, which for the
Hamiltonian of Eq.~(\ref{4}) yields, 

\begin{equation} \label{7}
m=\left\langle \left\langle \frac{\sinh (\widetilde{a}_{1}+\widetilde{a}%
_{2})}{\cosh (\widetilde{a}_{1}+\widetilde{a}_{2})+\exp (-2K_{1})\cosh (%
\widetilde{a}_{1}-\widetilde{a}_{2})}\right\rangle \right\rangle _{J^{(2)}}~, 
\end{equation}

\vskip \baselineskip
\noindent
where $\widetilde{a}_{1}=-\beta a_{1}$, $\widetilde{a}_{2}=-\beta a_{2}$, 
and $K_{1}=\beta J_{1}$.
Using the identity $\exp (aD_{x}+bD_{y})g(x,y)=g(x+a,y+b)$, where 
$D_{\mu }=\frac{\partial }{\partial \mu }$ ($\mu =x,y$) is the 
differential operator, Eq.~(\ref{7}) becomes

\begin{equation} \label{8}
m=\left\langle \left\langle \exp (\widetilde{a}_{1}D_{x}+\widetilde{a}%
_{2}D_{y})\right\rangle \right\rangle _{J^{(2)}}\left. 
g(x,y)\right\vert_{x,y=0}~, 
\end{equation}

\vskip \baselineskip
\noindent
with

\begin{equation} \label{9}
g(x,y)=\frac{\sinh (x+y)}{\cosh (x+y)+\exp (-2K_{1})\cosh (x-y)}~. 
\end{equation}

\vskip \baselineskip
\noindent
Applying the van der Waerden identity for the exponentials containing
Ising variables, i.e., 
$\exp (\alpha \sigma_{i})=\cosh(\alpha)+\sigma _{i}\sinh(\alpha)$ 
($\sigma _{i}=\pm 1$), 
the right-hand side of Eq.~(\ref{8}) can be 
written \textit{exactly} in terms of multiple spin correlation functions; 
however, it is clear that if one tries to
treat exactly all correlation functions, the problem
becomes unmanageable. In this work, we use a decoupling procedure that
ignores all high-order correlations on the right-hand side of 
Eq.~(\ref{8}), i.e., 

\begin{equation} \label{10}
\left\langle \left\langle \sigma _{i}\cdot \sigma _{j}\cdot \cdot \cdot
\sigma _{l}\cdot \sigma _{p}\right\rangle \right\rangle _{J^{(2)}}\simeq
\left\langle \left\langle \sigma _{i}\right\rangle 
\right\rangle _{J^{(2)}}\cdot
\left\langle \left\langle \sigma _{j}\right\rangle 
\right\rangle _{J^{(2)}}\cdot
\cdot \cdot \left\langle \left\langle \sigma _{l}\right\rangle \right\rangle
_{J^{(2)}}\cdot \left\langle \left\langle \sigma _{p}\right\rangle 
\right\rangle_{J^{(2)}}~, 
\end{equation}

\vskip \baselineskip
\noindent
where $i\neq j\neq \cdot \cdot \cdot \neq l\neq p$. The present procedure
neglects correlations between different spins, but takes into account 
relations such as $\left( \sigma _{i}\right) ^{2}=1$, whereas in the usual
mean-field approximation both self- and multi-spin correlation
functions are neglected.
Applying the approximation of Eq.~(\ref{10}) in Eq.~(\ref{8}), one gets the
equations of state,

\begin{equation} \label{11}
m=\sum\limits_{k=0}^{\tilde{k}}
A_{2k+1}^{(\Gamma)}(T,p)m^{2k+1}~, 
\end{equation}

\vskip \baselineskip
\noindent
where, due to the symmetry of the Hamiltonian, the even powers of $m$ do not
occur. In the equation above, the coefficients $A_{2k+1}^{(\Gamma)}(T,p)$
depend on the boundary
conditions shown in Fig.~\ref{fig1}, through the index 
$\Gamma={\rm F,SAF}$; in the present EFT-$2$ analysis 
one has that the highest power occurring on the right-hand side is defined
by $\tilde{k}=4$. 
In general, the coefficients $A_{2k+1}^{(\Gamma)}(T,p)$ depend on the size 
$n$ of the cluster used in EFT-$n$, getting more complicated 
for larger clusters; in addition to that, $\tilde{k}$ increases with $n$,
e.g., $\tilde{k}=3$ for $n=1$ and $\tilde{k}=5$ for $n=4$, in such a way
that the procedure may get very tedious for large clusters. 

The borders of continuous phase transitions between the ordered
({\bf F} and {\bf SAF}) phases and the high-temperature disordered  
({\bf P}) phase may be obtained by considering the limit  
$m\rightarrow 0$ in Eq.~(\ref{11}), which leads to 

\begin{equation} \label{12}
A_{1}^{(\Gamma)}(T_{c},p)=1~.  
\end{equation}
 
\begin{figure}[b]
\begin{center}
\includegraphics[width=0.49\textwidth,angle=0]{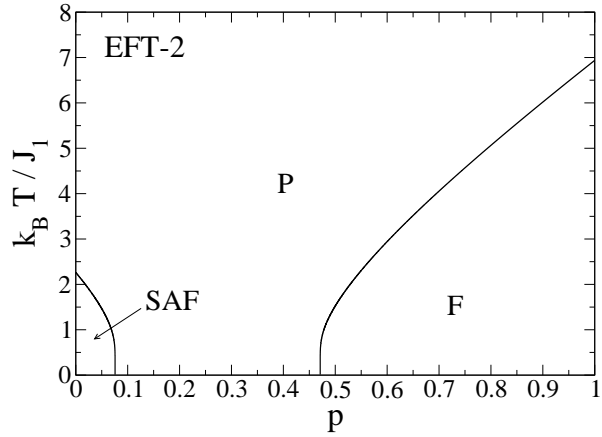}
\end{center}
\protect\caption{Phase diagram of the model defined in 
Eqs.~(\ref{1}) and (\ref{2}) for $J_{2}=J_{1}$ within EFT-2. 
The phases are the  
paramagnetic ({\bf P}), ferromagnetic ({\bf F}), and 
superantiferromagnetic ({\bf SAF}), as described in the text.}
\label{fig2}
\end{figure}

\vskip \baselineskip
\noindent
By solving Eq.~(\ref{12}) numerically, one finds $T_{c}(p)$ for both phases 
{\bf F} and {\bf SAF}, as shown in the phase diagram of 
Fig.~\ref{fig2}. 
For $p=0$ one has a continuous transition between the  
{\bf SAF} and {\bf P} phases at $(k_{B}T_{c}(0)/J_{1}) \cong 2.263$, 
which is in good agreement with the value 
$(k_{B}T_{c}(0)/J_{1}) \cong 2.083$ estimated from Monte Carlo simulations 
\cite{binderlandau80,malakis06}, as well as from a computation of 
the zeros of the partition function \cite{monroekim}. The critical
temperature associated with the {\bf SAF} phase decreases as 
$p$ increases, in
such a way that one gets $T_{c}(p_{c1})=0$, for $p_{c1} \cong 0.075$. Due to 
strong frustration effects, the system does not present long-range order 
in the interval $p_{c1}<p<p_{c2}$, where $p_{c2} \cong 0.471$. 
For $p>p_{c2}$ one gets
that the critical temperature associated with the {\bf F} phase
increases monotonically, leading to $(k_{B}T_{c}(1)/J_{1}) \cong 6.937$. 
Besides the results calculated within
EFT-2 shown in Fig.~\ref{fig2}, we have also computed these critical
frontiers within EFT-1 and 
EFT-4. Only in the EFT-1 case, the critical frontier {\bf SAF}-{\bf P}
comes as a first-order transition; we believe this to be a spurious
result (attributed to the smallness of the cluster), since it is in contrast
with those
obtained from EFT-2 and EFT-4 (which yield continuous phase transitions),
as well as with those of Monte Carlo simulations for the
particular case $p=0$ \cite{binderlandau80,malakis06}. From EFT-4 one gets
that $(k_{B}T_{c}(0)/J_{1}) \cong 2.310$ and $p_{c1} \cong 0.088$, 
indicating an enlargement
of the {\bf SAF} phase, whereas for the {\bf F} phase one has
$p_{c2} \cong 0.474$ and $(k_{B}T_{c}(1)/J_{1}) \cong 6.750$. 
It should be mentioned that our estimates for
$(k_{B}T_{c}(1)/J_{1})$, from EFT-1, EFT-2, and EFT-4, suggest a
slow convergence towards those from low- and high-temperature 
series expansions, which yield $(k_{B}T_{c}(1)/J_{1}) \cong 5.260$ and 
$(k_{B}T_{c}(1)/J_{1}) \cong 5.257$, respectively \cite{daltonwood}, or to
the more recent one, obtained from an analytical expression for the 
interface free energy, $(k_{B}T_{c}(1)/J_{1}) \cong 5.376$ \cite{zandvliet}.
In a similar way, our results suggest a decrease in the gap 
$(p_{c2}-p_{c1})$, although it is not possible to conclude whether the phases
{\bf SAF} and {\bf F} should meet at zero temperature.

It is important to mention that the EFT method, which replaces averages
over  products of spins by the respective products of their averages,
neglects correlations and, as a consequence, the associated critical
exponents are mean-field-like. Therefore, this method is not suitable for
investigations of possible changes in the critical exponents with respect
to variations of  important parameters of the problem  
(like $R$ and $p$), which represents a major question in its $p=0$
particular case, 
the $J_{1}-J_{2}$ Ising model. Furthermore, since the EFT-$n$ procedure
consists in treating a cluster of $n$ spins exactly, whereas the
interaction  
of this cluster with its surrounding are treated as an average 
(mean-field-like), the method becomes unmanageable for increasing values of
$n$. However, this technique presents the advantage of providing phase
diagrams that may be qualitatively correct as a whole, and in some cases it
may yield critical points with a good accuracy, in spite of  considering
clusters with small $n$, like in the following examples:  
(i) The tricritical point in the phase diagram of the $J_{1}-J_{2}$ Ising model 
on the square lattice, which was estimated to occur for 
$(J_{2}/J_{1})<1$ within EFT~\cite{rosana}, is in agreement 
with other methods, like Monte Carlo simulations~\cite{malakis06,kalz}
and analysis of zeros of the partition function~\cite{monroekim}; 
(ii) The present estimate of the critical temperature for $p=0$, 
 $(k_{B}T_{c}(0)/J_{1}) \cong 2.263$, yields a relative discrepancy 
 of typically $8 \%$ with respect to the value 
 $(k_{B}T_{c}(0)/J_{1}) \cong 2.083$, 
 estimated from Monte Carlo simulations 
\cite{binderlandau80,malakis06} and zeros of the partition function 
\cite{monroekim}.  
In the present problem, a full phase diagram was obtained for the case
$(J_{2}/J_{1})=1$, where an elimination of the {\bf SAF} phase occurs as one 
increases the concentration of second-neighbor ferromagnetic bonds.
Therefore, according to the present EFT approach, only for low values of
$p$ one gets a sufficient concentration of second-neighbor
antiferromagnetic bonds in order to yield the {\bf SAF} ordering at low
temperatures. As the concentration $p$ increases, the competition between
ferromagnetic couplings (both from the nearest-neighbor pairs and
second-neighbor pairs with probability $p$) and the second-neighbor  
antiferromagnetic ones [with probability $(1-p)$] destroys
the {\bf SAF} ordering at the concentration $p_{c1}$. Above this value, the
competition between ferro- and antiferromagnetic interactions leads to a
sufficient amount of frustrations that could, in principle, favor a  
type of spin-glass ordering at low temperatures. However, similarly to many
other frustrated two-dimensional systems, which do not exhibit a spin-glass
phase for finite temperatures, herein the paramagnetic phase dominates up
to zero temperature, in the interval $p_{c1}<p<p_{c2}$. For  
$p>p_{c2}$, one has sufficient ferromagnetic bonds leading to 
the {\bf F} ordering.  Obviously, the present estimates of $p_{c1}$ and 
$p_{c2}$ are expected to change under other approximation techniques 
(e.g., Monte Carlo simulations) and one cannot rule out the possibility of
a critical frontier separating the phases {\bf SAF} and {\bf F} at finite
temperatures.     

To conclude, we have introduced an Ising model characterized by 
ferromagnetic nearest-neighbor interactions $J_{1}$ and random  
$\pm J_{2}$ next-nearest-neighbor interactions [$+J_{2}$ 
with probability $p$ and $-J_{2}$ with probability $(1-p)$]. This model,
which generalizes the $J_{1}-J_{2}$ Ising model (herein corresponding to
the case $p=0$) is expected to be relevant for some diluted 
magnetic compounds, characterized
by a competition between nearest- and next-nearest-neighbor interactions. 
We have studied the phase diagram of this model in the particular case 
$J_{1}=J_{2}$ within the framework of an effective-field theory based on
the differential-operator technique, considering finite clusters with   
$n=1,2, {\rm and} \ 4$ spins.
We have found a low-temperature sublattice ordering associated with a
superantiferromagnetic phase for small $p$ ($0<p<p_{c1}$, where  
$p_{c1} \cong 0.088$ in the case $n=4$), as well as a ferromagnetic 
phase for larger values
of $p$ ($p>p_{c2}$, where $p_{c2} \cong 0.474$ in the case $n=4$), 
whereas in the interval 
$p_{c1}<p<p_{c2}$ the system does not present long-range order. 
Other approximation methods, such as renormalization-group and Monte Carlo
simulations, should be used to obtain further information on the critical
behavior of this model. 

\vskip 2\baselineskip

{\large\bf Acknowledgments}

\vskip \baselineskip
\noindent
J. R. S. thanks Centro Brasileiro de Pesquisas F\'{\i}sicas for the 
hospitability. This work was partially supported by CNPq, FAPEAM, FAPERJ, 
and CAPES (Brazilian Research Agencies). 

\vskip 2\baselineskip

\newpage

\begin{center}

{\large\bf Figure Captions}

\end{center}

\noindent
{\bf Fig. 1:} A cluster with $n=2$ (represented by spins   
$S_{1}$ and $S_{2}$) and its surrounding (spins 
$\sigma_{1},\sigma_{2}, \cdots, \sigma_{10}$), for the model defined in  
Eq.~(\ref{1}), in its ferromagnetic (a) and
superantiferromagnetic (b) ground states. The full lines represent the
interactions of spins $S_{1}$ and $S_{2}$, which are taken into account 
exactly in
Eq.~(\ref{3}), whereas the dashed lines represent the usual square-lattice
structure. 

\vskip \baselineskip
\noindent
{\bf Fig. 2:} Phase diagram of the model defined in 
Eqs.~(\ref{1}) and (\ref{2}) for $J_{2}=J_{1}$ within EFT-2. 
The phases are the  
paramagnetic ({\bf P}), ferromagnetic ({\bf F}), and 
superantiferromagnetic ({\bf SAF}), as described in the text. 

\newpage

\end{document}